\documentclass{elsart3p}        % Don't modify
\usepackage{graphicx}
\begin{document}
\begin{frontmatter}

\title{Magnetic and superconducting phase diagrams in ErNi$_2$B$_{2}$C}
\author[address1]{J.A Galvis, M. Crespo, I. Guillam\'on, H. Suderow\thanksref{thank1}, S. Vieira,}
\author[address2]{M. Garc\'ia Hern\'andez,}
\author[address3]{S. Bud'ko and P.C. Canfield}
\address[address1]{Laboratorio de Bajas Temperaturas, Departamento de
F\'isica de la Materia Condensada \\ Instituto de Ciencia de
Materiales Nicol\'as Cabrera, Facultad de Ciencias \\ Universidad
Aut\'onoma de Madrid, 28049 Madrid, Spain}
\address[address2]{Intituto de Ciencia de Materiales de Madrid,
 Consejo Superior de Investigaciones Cient\'ificas,
 Campus de Cantoblanco, E-28049 Madrid, Spain.}
\address[address3]{Ames Laboratory and Departament of Physics and Astronomy \\
Iowa State University, Ames, Iowa 50011, USA}

\thanks[thank1]{
* Corresponding author: hermann.suderow@uam.es}

\begin{abstract}
We present measurements of the superconducting upper critical field
H$_{c2}$(T) and the magnetic phase diagram of the superconductor
ErNi$_2$B$_2$C made with a scanning
tunneling microscope (STM). The magnetic field was applied in the
basal plane of the tetragonal crystal structure. We have found large gapless regions in the
superconducting phase diagram of ErNi$_2$B$_2$C, extending
between different magnetic transitions. A close
correlation between magnetic transitions and H$_{c2}$(T) is found, showing that superconductivity is strongly linked to magnetism.
\end{abstract}

\end{frontmatter}

ErNi$_2$B$_2$C is a prominent member of the series of the quaternary
rare earth nickel borocarbides (RNi$_2$B$_2$C with R=Gd-Lu and Y),
because of the coexistence of superconductivity with different kinds
of magnetic ordering in an accessible temperature
and magnetic field range\cite{Canfield98,Mueller01,Budko06}. It becomes superconducting
below about 11 K, goes into an incommensurate antiferromagnetic (AF) phase below
T$_N$=6.3 K, and develops a net weak ferromagnetic component (WF) below
T$_{WF}$=2.3 K, where one of each twenty spins of the initially
fully antiferromagnetic modulation are aligned parallel to each
other
\cite{Cava94b,Cho95,Zarestky95,Sinha95,Canfield96,Choi01,Kawano02}.
Remarkably, superconductivity survives all magnetic transitions,
contrary to what occurs in the Chevrel phase compounds (RRh$_4$B$_4$ with R being a rare earth), where a
reentrance to the normal phase is observed when entering a
ferromagnetic state\cite{FisherMaple,Bulaevski85,Kulic06,Prozorov08,Crespo09}. One of
the most salient features is a strong interrelationship between the
magnetic and superconducting phase diagrams, first observed in the
superconducting upper critical field\cite{Canfield96}. The magnetic
phase diagram has been studied using a variety of techniques.
Neutron scattering and magnetization data have revealed the form of
the spin alignment in a wide range of magnetic fields and
temperatures \cite{Jensen04,Budko00}. A magnetoelastic tetragonal to
orthorhombic distortion has been observed at T$_N$ using synchroton
X-ray scattering\cite{Detlefs97}, and the corresponding anomalies in
the thermal expansion have been followed in ref.\cite{Doerr02}. Thermal expansion, magnetostriction, and specific heat
measurements have revealed insight into the pressure dependence of
the transitions\cite{Budko06b}. Scanning Hall probe microscopy has
revealed the presence of a small random magnetic field all over the
surface of this compound \cite{Bluhm06}.

The superconducting
electronic density of states of ErNi$_2$B$_2$C, which is
unaccessible from specific heat due to the overwhelmingly large
magnetic contributions\cite{Budko06b,Massalami03}, has been studied
using scanning tunneling microscopy and spectroscopy. Instead of a
well developed BCS superconducting density of states a high amount
of excitations appear at the Fermi level, with a V-shaped density of
states, which persists all the way up to T$_c$ \cite{Crespo06a}, pointing towards a highly anomalous superconducting
density of states. TmNi$_2$B$_2$C orders
antiferromagnetically below 1.5 K, with a far smaller local magnetic
moment and has a well developed superconducting gap \cite{Suderow01}.
Pair breaking due to strong magnetic scattering, has been argued to be more important
in ErNi$_2$B$_2$C than in TmNi$_2$B$_2$C due to the stronger
exchange field in the former compound\cite{Gusakova06}. Recent thermal conductivity measurements in HoNi$_2$B$_2$C have shown that enhanced magnetic pair breaking exists in a large part of the phase diagram, with superconducting features in the thermal conductivity appearing below the resistively measured upper critical field\cite{Schneider09}.

\begin{figure}[]
\includegraphics[width=7.5cm,clip]{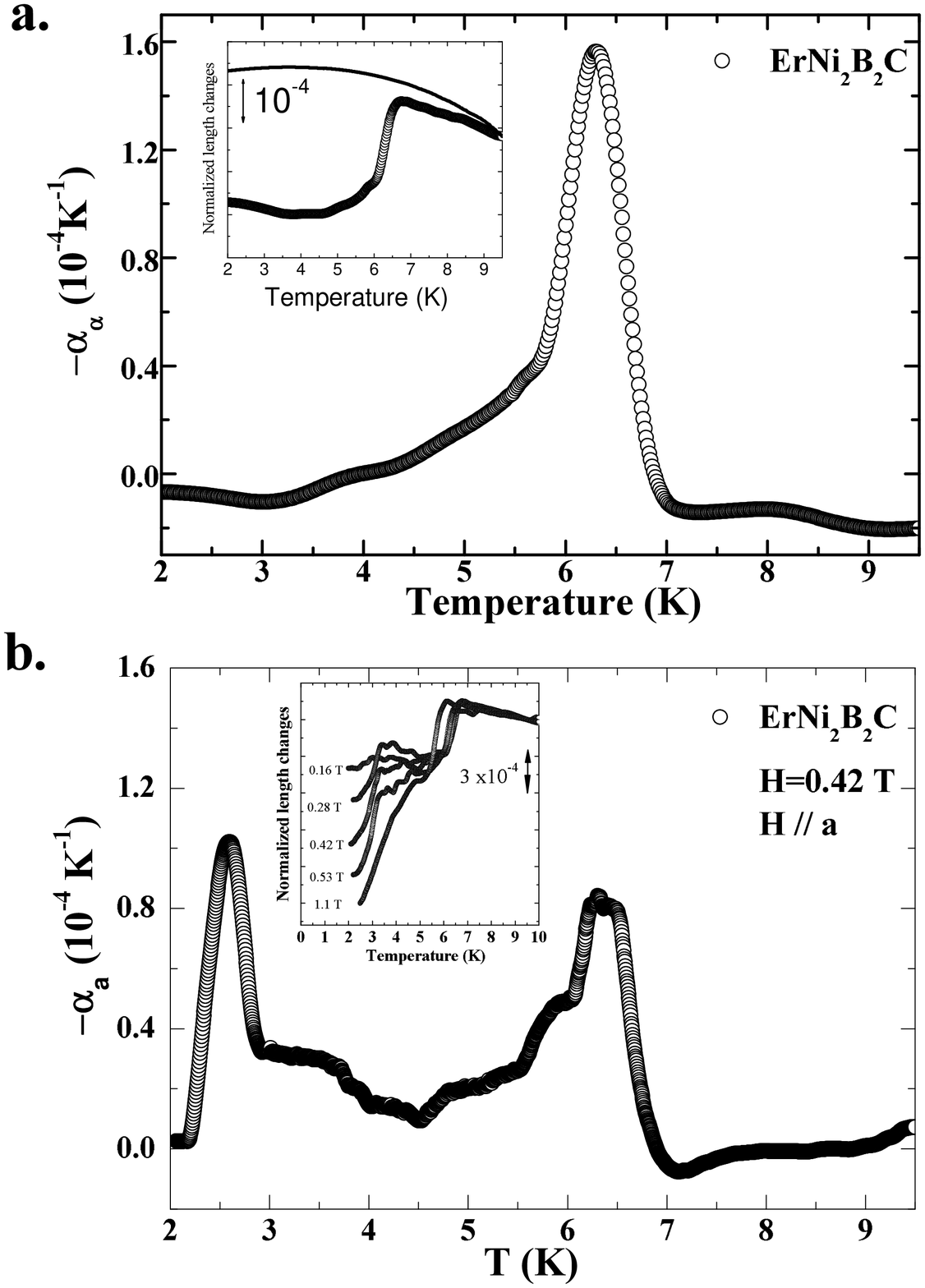}
\vskip -0cm
\caption{The longitudinal thermal expansion of the
sample perpendicular to the basal plane $\alpha_a$, as a function of
temperature at zero field (a) and under magnetic fields (b). In the
inset of (a) we show the changes in the z position of the tip,
together with the result obtained in a similar sample of
TmNi$_2$B$_2$C. In the inset of (b) we show a representative example
of the z position as a function of temperature at (from top to
bottom) 0.16 T, 0.28 T, 0.42 T, 0.53 T and 1.1 T.} \label{fig1}
\end{figure}

The objective of the present work is to measure H$_{c2}$(T) of ErNi$_2$B$_2$C using tunneling spectroscopy, and search for possible relationships
to the magnetic phase diagram. We have used similar experimental set-up and samples as in previous
works \cite{Crespo06a,Crespo06b,Suderow11b}, consisting of a STM in a dilution refrigerator. Here we use gold tips, and the
sample was prepared by cutting a suitable needle out of single
crystalline platelets of several mm thickness and one cm$^2$ cross
section oriented along the basal plane. The needles were broken in-situ, at low
temperatures, providing for clean surfaces. The magnetic field was applied parallel to
the tunneling direction, i.e. along the a axis. Topography and spectroscopic features have been discussed elsewhere\cite{Crespo06b,Guillamon10}. Identification of magnetic transitions with STM using the length changes produced by the thermal expansion of the sample was discussed previously in
Refs.\cite{Crespo06b}. We follow changes in the position of the tip over the surface of the sample of
ErNi$_2$B$_2$C, when varying the temperature\cite{Park09}. We observe strong changes close to magnetic transitions, related to the
longitudinal thermal expansion along the a axis. In Fig.\ref{fig1} we show anomalies at the magnetic
transitions, giving clear-cut peaks with a height comparable to results obtained with other
methods\cite{Budko06b,Doerr02}. The position of the peaks give the
magnetic transition temperatures, which are comparable to previous work.

\begin{figure}[]
\includegraphics[width=7.5cm,clip]{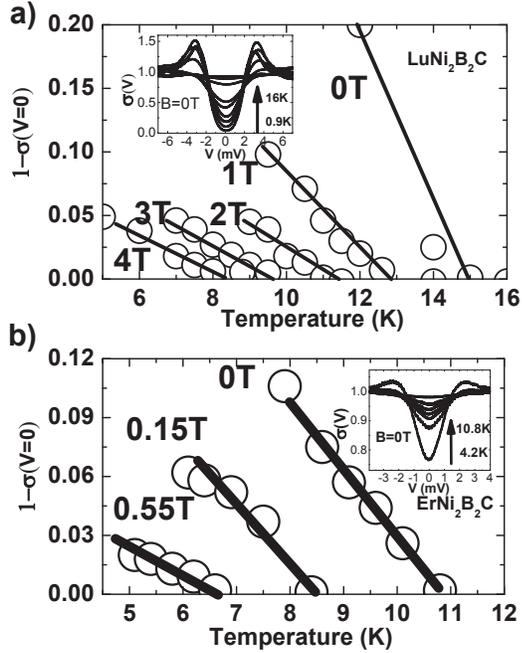}
\vskip -1cm
\caption{We show 1-$\sigma$(V=0) ($\sigma$(V=0) is the normalized zero bias conductance) as a function of the temperature for LuNi$_2$B$_2$C (a) and for ErNi$_2$B$_2$C (b) at several magnetic fields. The insets show corresponding tunneling spectroscopy curves. The size of the points is about 1\% of the high bias conductance and give experimental uncertainty. Lines are guides to the eye. The extrapolation of the lines to the x-axis gives the critical temperature at the given magnetic fields. Similar curves are obtained when varying the magnetic field at a fixed temperature.} \label{fig2}
\end{figure}

\begin{figure}[]
\includegraphics[width=7.5cm,clip]{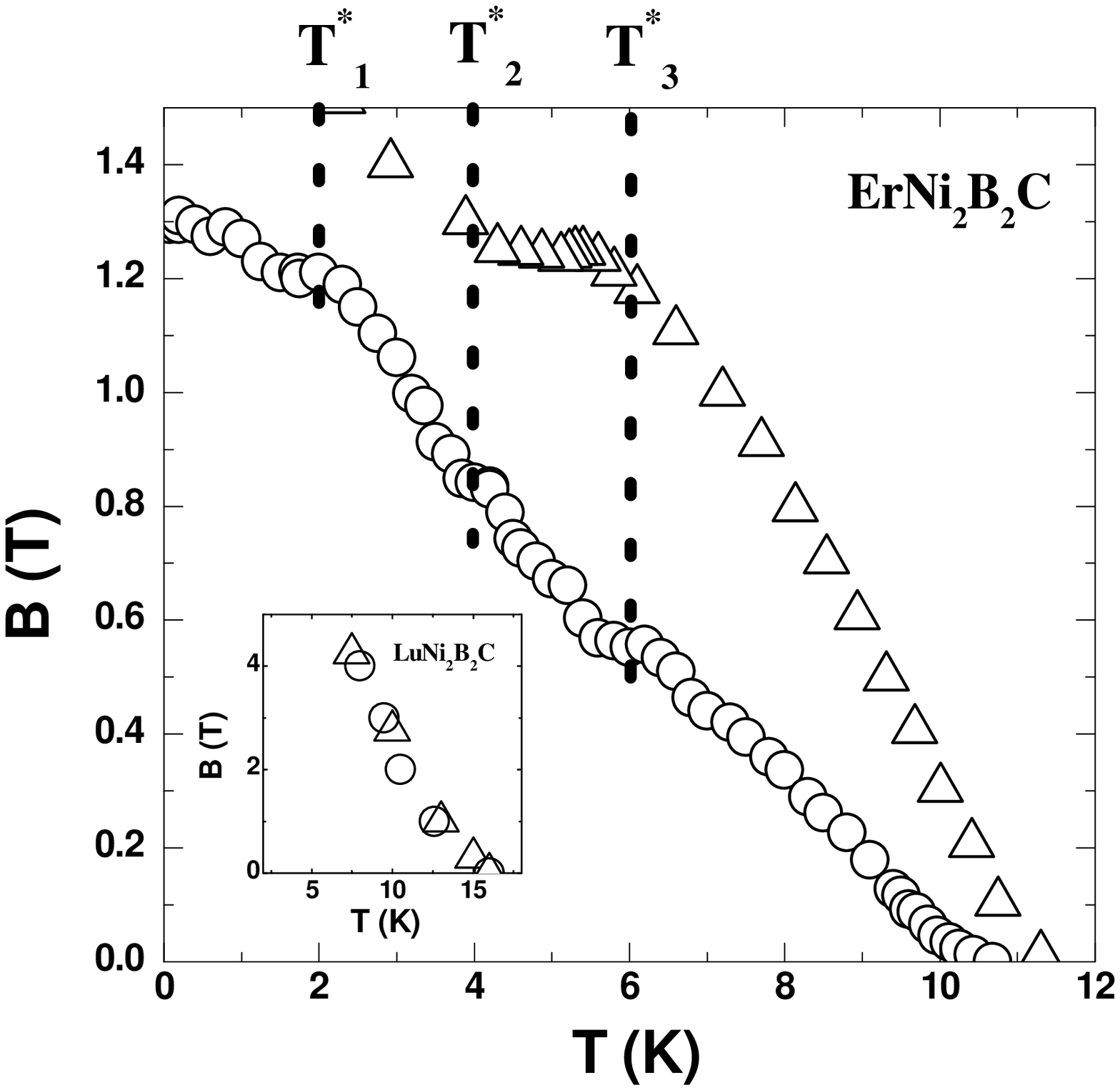}
\vskip -3cm
\caption{The superconducting phase diagram of
ErNi$_2$B$_2$C, down to 100 mK, with the field applied within the
basal plane, obtained from local tunneling spectroscopy (open points) and from resistivity (triangles, taken from Ref.\protect\cite{Jensen04}).
Note that Features in H$_{c2}$(T) are observed at approximately
T$_1^* \approx 2 K$, T$_2^*  \approx 4 K$ and T$_3^* \approx 6 K$
(dashed lines). Inset shows the same data (circles tunneling spectroscopy, triangles resistivity) in LuNi$_2$B$_2$C.} \label{fig3}
\end{figure}

To measure the upper critical field, we follow the superconducting tunneling features as a function of
temperature and magnetic field, in different positions on the surface\cite{Guillamon09}. The superconducting features are weak in ErNi$_2$B$_2$C, so we did not resolve the vortex lattice\cite{Guillamon10}.
We trace the temperature of field dependence of the zero bias conductance, normalized to its value at high bias. In the insets of Fig.2 we show some representative tunneling conductance curves in ErNi$_2$B$_2$C and in the non-magnetic parent compound LuNi$_2$B$_2$C. We define H$_{c2}$(T) as the point where the normalized zero bias conductance (Fig.2) reaches 0.99 when increasing temperature or magnetic field. At this point, the tunneling curves are essentially flat and voltage independent below some 10 mV, within the experimental uncertainty. In the Fig.\ref{fig3} we show H$_{c2}$(T) thus obtained. Note that although the data taken in LuNi$_2$B$_2$C closely follow previous results obtained by tracing the resistive transition (Fig.3), there is a large difference between H$_{c2}$(T) found in ErNi$_2$B$_2$C through tunneling spectroscopy and the result obtained by the resistive transition.
This means that there is a large temperature and field range where ErNi$_2$B$_2$C is essentially featureless in the tunneling conductance, still showing zero resistance. In TmNi$_2$B$_2$C, which has a transition to an antiferromagnetic state below 1.5 K, and where the gap is fully open at low temperatures\cite{Guillamon10,Suderow01,Eskildsen98,Norgaard00,Norgaard04}, we did not find differences between tunneling data and the resistive transition. This would imply that magnetism of
ErNi$_2$B$_2$C leads to a significant destruction of the superconducting
gap in applied magnetic fields.

Note on the other hand that data in ErNi$_2$B$_2$C have been taken down to very low temperatures, extending the range studied in previous work \cite{Canfield96,Budko00,Budko06}. There is a
significant positive curvature at high temperatures, characteristic of many nickel borocarbide superconductors.
This strong positive curvature has been associated with the
anisotropy of the Fermi surface \cite{Shulga98,Brison04,Suderow04b,Suderow05d}. There are
features at T$_3^* \approx 6$ K, T$_2^* \approx 4$ K and T$_1^* \approx 2$ K.
The features at 6 K and 2 K are pronounced, and H$_{c2}$(T)
becomes nearly flat at these points.

\begin{figure}
\includegraphics[width=7.5cm,clip]{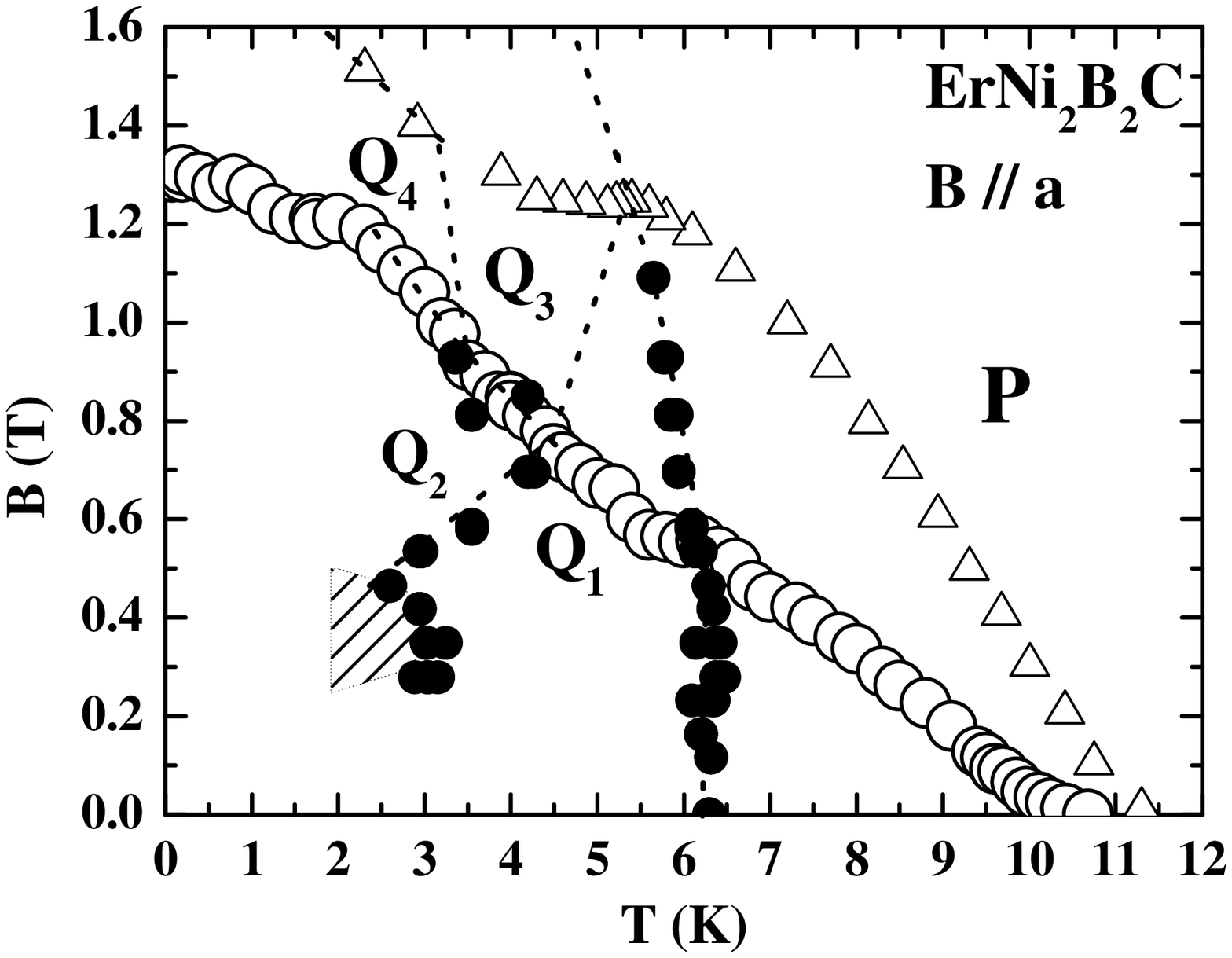}
\vskip -4cm
\caption{The magnetic and superconducting phase diagrams
of ErNi$_2$B$_2$C with the field applied within the basal plane.
Closed circles are points of the magnetic phase diagram obtained by
thermal expansion (see also
refs.\protect\cite{Doerr02,Budko06}), and open circles the
H$_{c2}$(T) data obtained by tunneling spectroscopy (same as
Fig.\protect\ref{fig2}). Open triangles are H$_{c2}$(T) data
obtained through resistivity, from ref.\protect\cite{Jensen04}.
Dashed lines are the magnetic transitions viewed in neutron
scattering experiments (from ref.\protect\cite{Jensen04}), and the
different magnetic phases are labeled by
Q$_1$-Q$_4$\protect\cite{Jensen04}. The magnetic states are labeled with P
for paramagnetic state, and with the corresponding Q vector,
determined in ref.\cite{Jensen04}, for the magnetic states.
} \label{fig4}
\end{figure}

In Fig.\ref{fig4} we compare the magnetic and superconducting phase
diagrams in ErNi$_2$B$_2$C.  As shown in Ref.\cite{Jensen04}, when a magnetic field is applied in plane between T$_{WF}$
and T$_{N}$, the initial incommensurate antiferromagnetic modulation
(Q=0.55-0.56 a$^*$), is modified into a field induced weak
ferromagnetic state which locks to the lattice and becomes
commensurate with a wavevector Q=(m/n)a$^*$, being m odd and n even
in most of the phases discussed in ref.\cite{Jensen04}. When the
magnetic field is increased, the wavelength of the commensurate
structures is gradually reduced, and the number of spins aligned
with the magnetic field increases. This produces new commensurate
phases identified as well in neutron scattering measurements as in
magnetization, which shows a steep increase at each
transition\cite{Canfield96,Budko00}.  The phases labeled in the
Fig.\ref{fig4} have been identified as the low field incommensurate
AF phase Q$_1$=0.55-0.56 a$^*$, and the commensurate phases
Q$_2$=0.57 a$^*$, Q$_3$=0.58 a$^*$ and Q$_4$=0.59 a$^*$. The final
transition to the saturated paramagnetic state occurs at about 2 T
along the a axis, for fields much above the loss of detectable superconducting gap via tunneling
spectroscopy. The features at H$_{c2}$(T) curve measured using tunneling spectroscopy at T$_2^*$ (Fig. \ref{fig3}) actually coincides with a
triple point (transition Q$_1$-Q$_2$-Q$_3$) in the magnetic phase
diagram. The rest of the H$_{c2}$(T) curve obtained using tunneling spectroscopy follows the magnetic transitions
Q$_2$-Q$_3$ and Q$_2$-Q$_4$ from 4.5 K to 2 K. Below about 2 K, another feature with a vanishing slope
 is observed in H$_{c2}$(T) curve obtained using tunneling spectroscopy. This possibly coincides with another
magnetic transition, evolving from the low temperature weakly
ferromagnetic state which exists at zero field. Note that the extended gapless region with zero resistance (Fig.4) is bound within the magnetic transitions Q$_2$-Q$_4$, showing that the density of states in this compound is significantly influenced by the underlying magnetic state. Thus, possible explanations for the gapless state can be enhanced magnetic scattering near the surface\cite{Gusakova06}, or band-dependent strong pair breaking\cite{Schneider09}.

The strong correlation between magnetism and superconductivity in ErNi$_2$B$_2$C leads to remarkable features in the superconducting phase diagram, which closely follows the changes in the local spin arrangement. The question of how these are linked together remains open for further study. Measurements of the vortex lattice, or of the tunneling spectroscopy using superconducting tips\cite{Rodrigo04b,Suderow09}, should be useful, as well as other macroscopic measurements, such as thermal conductivity or London penetration depth.

\section{Acknowledgments.}

The Laboratorio de Bajas Temperaturas is associated to the ICMM of the CSIC. This work was supported by the U.S. Department of Energy, Office of Basic Energy Science, Division of Materials Sciences and Engineering. Part of this work was performed at the Ames Laboratory. Ames Laboratory is operated for the U.S. Department of Energy by Iowa State University under Contract No. DE-AC02-07CH11358. This work was also
supported by the spanish MICINN (grant FIS2011-23488, ACI-2009-090 and Consolider Ingenio 2010 Nanociencia Molecular) and the Comunidad de Madrid through program Nanobiomagnet.

%\bibliography{Lastbib5}

\begin{thebibliography}{10}
\expandafter\ifx\csname url\endcsname\relax
  \def\url#1{\texttt{#1}}\fi
\expandafter\ifx\csname urlprefix\endcsname\relax\def\urlprefix{URL }\fi

\bibitem{Canfield98}
P.~C. Canfield, P.~L. Gammel, D.~J. Bishop, New magnetic superconductors: A toy
  box for solid-state physicists, Physics Today 51 (1998) 40.

\bibitem{Mueller01}
K.~H. M\"{u}ller, V.~Narozhnyi, Interaction of superconductivity and magnetism
  in borocarbide superconductors, Reports Prog. Phys. 64 (2001) 943.

\bibitem{Budko06}
S.~Bud'ko, P.~Canfield, Magnetism and superconductivity in the rare-earth
  nickel borocarbides, C.R. Physique 7 (2006) 56.

\bibitem{Cava94b}
R.~Cava, H.~Takagi, , B.~Batlogg, H.~Zandbergen, J.~Krajewski, W.~Peck, R.~van
  Dover, R.~Felder, T.~Siegrist, K.~Mizuhashi, J.~Lee, H.~Eisaki, S.~Uchida,
  Superconductivity in the quaternary intermetallic compounds $lnni_{2}b_{2}c$,
  Nature 367 (1994) 252.

\bibitem{Cho95}
B.~K. Cho, P.~C. Canfield, L.~L. Miller, D.~C. Johnston, W.~P. Beyermann,
  A.~Yatskar, Magnetism and superconductivity in single-crystal
  $erni_{2}b_{2}c$, Phys. Rev. B 52 (1995) 3684.

\bibitem{Zarestky95}
J.~Zarestky, C.~Stassis, A.~I. Goldman, P.~C. Canfield, P.~Dervenagas, B.~K.
  Cho, D.~C. Johnston, Magnetic structure of $erni_{2}b_{2}c$., Physical Review
  B 51 (1995) R678.

\bibitem{Sinha95}
S.~Sinha, J.~Lynn, T.~Grigereit, Z.~Hossain, L.~Gupta, R.~Nagarajan, C.~Godart,
  Neutron-diffraction study of antiferromagnetic order in the magnetic
  superconductor erni2b2c, Phys. Rev. B 51 (1995) 681.

\bibitem{Canfield96}
P.~C. Canfield, S.~L. Bud'ko, B.~K. Cho, Possible co-existence of
  superconductivity and weak ferromagnetism in $erni_{2}b_{2}c$., Physica C 262
  (1996) 249.

\bibitem{Choi01}
S.~M. Choi, J.~W. Lynn, D.~Lopez, P.~L. Gammel, P.~C. Canfield, S.~L. Bud'ko,
  Direct observation of spontaneous weak ferromagnetism in the superconductor
  $erni_{2}b_{2}c$, Phys. Rev. Lett. 87 (2001) 107001.

\bibitem{Kawano02}
H.~Kawano-Furukawa, H.~Takeshita, M.~Ochiai, T.~Nagata, H.~Yoshizawa,
  N.~Furukawa, H.~Takeya, K.~Kadowaki, Weak ferromagnetic order in the
  superconducting $erni_{2}^{11}b_{2}c$, Phys. Rev. B 65 (2002) 180508(R).

\bibitem{FisherMaple}
E.~O. Fisher, M.~Maple, Superconductivity in ternary compounds, Vols I and II,
  Springer, New York.

\bibitem{Bulaevski85}
L.~N. Bulaevski, A.~I. Buzdin, M.~L. Kulic, S.~V. Panjukov, Coexistence of
  superconductivity and magnetism: Theoretical predictions and experimental
  results, Advances in Physics 34 (1985) 175 -- 261.

\bibitem{Kulic06}
M.~Kulic, Conventional magnetic superconductors: coexistence of singlet
  superconductivity and magnetic order, C.R. Physique 7 (2006) 4.

\bibitem{Prozorov08}
R.~Prozorov, M.~Vannette, S.~Law, S.~Bud'ko, P.~Canfield, Physical Review B 77
  (2008) 100503.

\bibitem{Crespo09}
V.~Crespo, J.~Rodrigo, H.~Suderow, S.~Vieira, D.~Hinks, I.~Schuller, Phys. Rev.
  Lett. 102 (2009) 237002.

\bibitem{Jensen04}
A.~Jensen, K.~N. Toft, A.~B. Abrahamsen, D.~F. McMorrow, M.~R. Eskildsen, N.~H.
  Andersen, J.~Jensen, P.~Hedegard, J.~Klenke, S.~Danilkin, K.~Probes,
  V.~Sikolenko, P.~Smeibidl, S.~L. Bud'ko, P.~C. Canfield, Field-induced
  magnetic phases in the normal and superconducting states of erni$_2$b$_2$c,
  Phys. Rev. B 69 (2004) 104527.

\bibitem{Budko00}
S.~Bud'ko, P.~Canfield, Rotational tuning of $H_{c2}$ anomalies in
  $erni_{2}b_{2}c$: Angular-dependent superzone gap formation and its effect on
  the superconducting ground state, Phys. Rev. B. 61 (2000) R14932.

\bibitem{Detlefs97}
C.~Detlefs, A.~H. M.~Z. Islam, T.~Gu, A.~I. Goldman, C.~Stassis, P.~C.
  Canfield, Magnetoelastic tetragonal-to-orthorhombic distortion in
  $erni_{2}b_{2}$c, Phys. Rev. B. 56 (1997) 7843.

\bibitem{Doerr02}
M.~Doerr, M.~Rotter, M.~Massalami, S.~Sinning, H.~Takeya, Magnetoelastic
  effects in $erni_{2}b_{2}c$ single crystal: probing the h-t phase diagram, J.
  Phys.: Condens. Matter 14 (2002) 5609.

\bibitem{Budko06b}
S.~Bud'ko, G.~Schmiedeshoff, P.~Canfield, Anisotropic thermal expansion and
  uniaxial pressure dependence of superconducting and magnetic transitions in
  erni$_2$b$_2$c, Solid State Comm. 140 (2006) 281.

\bibitem{Bluhm06}
H.~Bluhm, S.~Sebastian, J.~Guikema, I.~Fisher, K.~Moler, Scanning hall probe
  imaging of erni2b2c, Physical Review B 73 (2006) 014514.

\bibitem{Massalami03}
M.~E. Massalami, R.~E. Rapp, F.~A.~B. Chaves, H.~Takeya, C.~M. Chaves, Magnon
  specific heat of single-crystal borocarbides $rni_{2}b_{2}c$ (r = tm, er, ho,
  dy, tb, gd), Phys. Rev. B 67 (2003) 224407.

\bibitem{Crespo06a}
M.~Crespo, H.~Suderow, S.~Bud'ko, P.~C. Canfield, S.~Vieira, Local
  superconducting density of states of {ErNi$_{2}$B$_{2}$C}, Phys. Rev. Lett.
  96 (2006) 027003.

\bibitem{Suderow01}
H.~Suderow, P.~Mart\'inez-Samper, S.~Vieira, N.~Luchier, J.~P. Brison, P.~C.
  Canfield, Tunneling spectroscopy in the magnetic superconductor
  {TmNi$_{2}$B$_{2}$C}, Phys. Rev. B 64 (2001) 020503(R).

\bibitem{Gusakova06}
D.~Gusakova, A.~Golubov, M.~Kupriyanov, A.~Buzdin, Density of states in sf
  bilayers with arbitrary strength of magnetic scattering, Pis'ma V ZhETF 83
  (2006) 385.

\bibitem{Schneider09}
M.~Schneider, G.~Fuchs, K.-H. Muller, K.~Nenkov, G.~Behr, D.~Souptel,
  S.~Drechsler, Physical Review B 80 (2009) 224522.

\bibitem{Crespo06b}
M.~Crespo, H.~Suderow, S.~Vieira, S.~Bud'ko, P.~C. Canfield, Thermal expansion
  measured by stm in the magnetic superconductor $erni_{2}b_{2}$c, Physica B
  378 (2006) 471.

\bibitem{Suderow11b}
H.~Suderow, I.~Guillamon, S.~Vieira, Rev. of Sci. Inst. 82 (2011) 033711.

\bibitem{Guillamon10}
I.~Guillamon, M.~Crespo, H.~Suderow, S.~Vieira, J.~Brison, S.~Bud'ko,
  P.~Canfield, Physica C 470 (2010) 771--775.

\bibitem{Park09}
J.-H. Park, D.~Graf, T.~Murphy, G.~Schmiedeshoff, S.~Tozer, Rev. Sci. Inst. 80
  (2009) 116101.

\bibitem{Guillamon09}
I.~Guillamon, H.~Suderow, A.~Fernandez-Pacheco, J.~Sese, R.~Cordoba, J.~D.
  Teresa, M.~Ibarra, S.~Vieira, Nature Physics 5 (2009) 651.

\bibitem{Eskildsen98}
M.~R. Eskildsen, K.~Harada, P.~L. Gammel, A.~B. Abrahamsen, N.~H. Andersen,
  G.~Ernst, A.~P. Ramirez, D.~J. Bishop, K.~Mortensen, D.~G. Naugle, K.~D.~D.
  Rathnayaka, P.~Canfield, Intertwined symmetry of the magnetic modulation and
  the flux-line lattice in the superconducting state of {TmNi$_{2}$B$_{2}$C},
  Nature 393 (1998) 242--245.

\bibitem{Norgaard00}
K.~Noorgaard, M.~R. Eskildsen, N.~H. Andersen, Interdependence of magnetism
  and superconductivity in the borocarbide {$TmNi_{2}B_{2}C$}, Phys. Rev. Lett.
  84 (2000) 4982.

\bibitem{Norgaard04}
K.~Noorgaard, A.~B. Abrahamsen, M.~R. Eskildsen, K.~Lefmann, N.~H. Andersen,
  P.~Vorderwisch, P.~Smeibidl, M.~Meissner, P.~C. Canfield, Neutron diffraction
  study of anomalous high-field magnetic phases in {TmNi$_{2}$B$_{2}$C}, Phys.
  Rev. B. 69 (2004) 214507.

\bibitem{Shulga98}
S.~V. Shulga, S.-L. Drechsler, G.~Fuchs, K.~H. M\"{u}ller, K.~Winzer,
  M.~Heinecke, K.~Krug, Upper critical field peculiarities of superconducting
  $yni_{2}b_{2}c$ and $luni_{2}b_{2}c$, Phys. Rev. Lett. 80 (1998) 1730.

\bibitem{Brison04}
J.~Brison, N.~Luchier, A.~Sulpice, H.~Suderow, P.~Mart\'inez-Samper, S.~Vieira,
  A.~Buzdin, P.~Canfield, Anisotropic superconductivity in borocarbide
  superconductors and spin disorder, Journal of Magnetism and Magnetic
  Materials 272 (2004) 158.

\bibitem{Suderow04b}
H.~Suderow, V.~G. Tissen, J.~P. Brison, J.~L. Martinez, S.~Vieira, P.~Lejay,
  S.~Lee, S.~Tajima, Pressure dependence of the upper critical field of mgb$_2$
  and of yni$_2$b$_2$c, Phys. Rev. B 70 (2004) 134518.

\bibitem{Suderow05d}
H.~Suderow, V.~Tissen, J.~Brison, J.~Martinez, S.~Vieira, Pressure induced
  effects on the {Fermi} surface of superconducting {2H-NbSe$_2$}, Phys. Rev.
  Lett. 95 (2005) 117006.

\bibitem{Rodrigo04b}
J.~G. Rodrigo, H.~Suderow, S.~Vieira, E.~Bascones, F.~Guinea, Superconducting
  nanostructures fabricated with the scanning tunnelling microscope, J. Phys.:
  Condens. Matter 16 (2004) 1151.

\bibitem{Suderow09}
H.~Suderow, V.~Crespo, I.~Guillamon, S.~Vieira, F.~Servant, P.~Lejay,
  J.~Brison, J.~Flouquet, A nodeless superconducting gap in {Sr$_2$RuO$_4 $}
  from tunneling spectroscopy, New Journal of Physics 11 (2009) 093004.

\end{thebibliography}

\end{document}